\documentclass[10pt,aps,amsmath,latexsym,amsfonts]{article}
\pdfoutput=1 

\usepackage{epsfig}
\usepackage{amsfonts,amssymb,amsmath}
\usepackage[hang]{subfigure}

\usepackage[T1]{fontenc} 

\author{Gianmassimo,\,Kazuya\thanks{Email: gianmassimo.tasinato@port.ac.uk}
\\
\it{Institute of Cosmology \& Gravitation, University of Portsmouth,
PO1 3FX, UK \\}}

\newcommand{\be}{\begin{equation}}
\newcommand{\ee}{\end{equation}}

\newcommand{\bi}{\begin{itemize}}
\newcommand{\ei}{\end{itemize}}

\input epsf.sty
\newcommand{\bea}{\begin{eqnarray}}
\newcommand{\eea}{\end{eqnarray}}

\newcommand{\bei}{\begin{itemize}}
\newcommand{\eei}{\end{itemize}}

\newcommand{\bright}{\begin{flushright}}
\newcommand{\eright}{\end{flushright}}
\newcommand{\bminip}{\begin{minipage}}
\newcommand{\eminip}{\end{minipage}}
\newcommand{\bcent}{\begin{center}}
\newcommand{\ecent}{\end{center}}

\def\tr{{\rm tr}}

\begin{document}





\begin{center}
{\LARGE \bf
The role of vector fields in modified gravity scenarios
}\\[1cm]

{
{\bf
Gianmassimo Tasinato$^{a\,}$\,,\, Kazuya Koyama$^{a\,}$ \,,\, Nima Khosravi$^{b\,c\,d\,}$
}
}
\\[7mm]
{\it $^a$
 Institute of Cosmology and Gravitation, University of Portsmouth,
 Portsmouth,
PO1 3FX, United Kingdom,} 
\\[3mm]
{\it $^b$
Cosmology Group, African Institute for Mathematical
Sciences, Muizenberg, 7945, South Africa}
\\[3mm]
{\it $^c$
South
African Astronomical Observatory, Observatory Road, Observatory,
Cape Town, 7935, South Africa
}\\[3mm]
{\it $^d$
Department of
Mathematics and Applied Mathematics, University of Cape Town,
Rondebosch, Cape Town, 7700, South Africa
}
\\[1cm]
\vspace{-0.3cm}

\vspace{1cm}

{\large\bf Abstract}

\end{center}
\begin{quote}

{
Gravitational vector degrees of freedom typically arise in many examples of modified gravity models.
We start to systematically explore their role in these scenarios,
studying the effects of coupling gravitational vector and scalar degrees of freedom. We
focus on set-ups that enjoy a Galilean symmetry in the scalar sector and an Abelian gauge symmetry 
in the vector sector. These symmetries, together with the requirement that the equations of motion 
contain at most two space-time derivatives, only allow for a small number of operators
in the Lagrangian for the gravitational fields. We investigate
the role of gravitational vector fields for two broad classes of phenomena that characterize modified gravity scenarios.
The first is self-acceleration: we analyze in general terms the behavior of vector fluctuations
around self-accelerating solutions,  and show that  vanishing kinetic terms of vector fluctuations  
lead  to instabilities on cosmological backgrounds.
  The second phenomenon is the screening of long range fifth forces by means of Vainshtein mechanism. We show
that if gravitational vector fields are appropriately coupled to a  spherically symmetric source,
they can play an important role for defining the features of the  background solution and the scale of the Vainshtein 
radius. Our general results can be applied to any concrete model of modified gravity, whose low-energy
vector and scalar degrees of freedom satisfy the symmetry  requirements that we impose.
}

\end{quote}

\section{Introduction}

General relativity  is the unique theory describing the dynamics of
an interacting spin-2, massless degree of freedom. Any consistent
modification of general relativity (GR) introduces new light
dynamical degrees of freedom  (DOFs) of lower spins, typically scalars and vectors,
which together with tensors constitute the low energy spectrum of  modified gravity theories.
These additional fields can change the strength of gravitational
interactions, for example rendering gravity weaker at large scales,
and provide an intriguing  explanation for  the observed present day
acceleration of our universe. Moreover, theories of modified gravity
are characterized by interesting environmental effects, since the
strength of  gravitational interactions depends on the particular
background under consideration. For example nearby a  massive source as the sun,
screening mechanisms are  capable to hide the  additional DOFs, recovering with excellent
accuracy the predictions of GR, and  satisfying stringent solar
system constraints on deviations from it. For a recent comprehensive
review on modified gravity see \cite{Clifton:2011jh}.

 \smallskip
Usually, a low energy effective description of modified gravity scenarios is made in terms
of a Lagrangian controlling  the dynamics of tensor and scalar modes. The dynamics of the
scalar, in particular, is assumed to control with good approximation the effects of the
modification of gravity. In this work we start to systematically explore the  role of
vectors in modified gravity, specifically studying the effects of  couplings between
vector and scalar DOFs. We take the point of view that vector DOFs are part of the set
of fields that control gravitational interactions.  Typically, they arise as Goldstone
bosons associated with symmetry breaking effects that characterize modified  gravity scenarios.
Representative examples are brane-world models, in which translational invariance in the extra
dimensions is broken by the presence of branes; and massive gravity, in which the graviton mass
breaks the diffeomorphism invariance of GR. In both of these cases, vectors appear as Goldstone
bosons of broken symmetries. Moreover, we will also comment on a second, alternative perspective,
in which vectors  are  part of the matter sector of the theory  under consideration, and couple
by means of  higher order operators with the scalar DOF that mediate gravitational interactions
in modified gravity.

\smallskip

Whatever the perspective we adopt, we are generally allowed to write a large
number of effective operators that couple vectors with the gravitational scalar  DOF.
This number can be reduced by imposing appropriate symmetries and physical constraints on the theory.
In this work, we  consider two symmetries: a Galilean symmetry in the scalar sector, and an Abelian
gauge symmetry in the vector sector. The Galilean symmetry demands that the action is invariant under
$\pi\,\to\,\pi+c+b_\mu x^\mu$, where $\pi$ is the gravitational scalar DOF. The $U(1)$ Abelian gauge symmetry
in the vector sector reads $A_\mu\,\to\,A_\mu+\partial_\mu \xi$. Such a gauge symmetry is usually encountered
in vector Lagrangians arising from symmetry breaking effects (for example, in massive gravity
 it
is a remnant of
broken diffeomorphism invariance  \cite{ArkaniHamed:2002sp}). As a physical requirement, we demand that our
Lagrangian leads to equations of motion with at most two derivatives, so to automatically avoid
  Ostrogradsky  instabilities. The theory we analyze is a generalization of scalar Galileons
to the so-called $p$-form Galileons
 \cite{Deffayet:2010zh,Tasinato:2012ze}. 
Notice that our work on coupling vectors to scalars by means of derivative interactions, such  to lead to 
equations of motion with at most two space-time derivatives, is complementary to the work
of Hordenski \cite{Horndeski:1976gi}, that found the most general vector-tensor Lagrangian satisfying the same requirement.

\smallskip

We will focus on investigating the role of vector fields for two broad classes of phenomena that characterize
the most interesting versions of modified gravity scenarios. The first is self-acceleration: theories with
higher order derivative scalar self-interactions admit cosmological solutions
corresponding to accelerating universes in the vacuum. The most famous example is the DGP model \cite{Dvali:2000hr}
whose basic features have then been extended in \cite{Nicolis:2008in} to ghost-free set-ups
enjoying a Galilean symmetry. While the contributions of the scalar sector to self-acceleration is well understood,
much less studied is the role of a vector sector derivatively coupled to the scalar.
In the set-up we consider, vectors can  acquire a non-trivial profile that preserves the
symmetries of cosmological backgrounds (i.e. the
Friedmann-Robertson-Walker symmetry of the metric) and, at the same
time, contributes to determine the size of cosmological acceleration.
Moreover, the behavior of vector perturbations around self-accelerating configurations can provide key
constraints to characterize the stability of a given cosmological background. In our
work, we will analyze in general terms possible instabilities that vectors induce on cosmological solutions
in Galileon models, and provide criteria to avoid them.

The second topic we will consider is the role of vectors for characterizing screening mechanisms, in particular
the Vainshtein effect.  In Galileon theories, within a certain distance from a spherically symmetric source,
the predictions of GR can be recovered despite the presence of additional light degrees of freedom besides
 tensors, thanks to the non-linear contributions to the equations of motion.
Usually, in these theories one considers minimal couplings between tensor and scalar gravitational DOFs to the source.
On the other hand, gravitational vector degrees of freedom are also allowed to couple { directly} to the source without violating the symmetries we impose on the theory. This happens if the source is characterized,
besides its usual energy momentum tensor, by a gravitational vector current.
In this case the source has  a gravitational vector charge, and we will show that this fact can
considerably influence the realization of the Vainshtein mechanism.
We will also comment   on how our findings can  be applied to set-ups in which
 scalars  enjoying   Galileon symmetries  couple to bodies  that are
 charged under electromagnetic interactions. Also in this case the realization of the
 Vainhstein mechanism might be influenced  by the vector charges.


\smallskip

This work is organized as follows. In Section \ref{sec-lagrangian} we present the Lagrangian
for the theory we consider. In Section \ref{section-max} we consider
 its maximally symmetric solutions in the vacuum, and analyze the role
 of vector fields for the stability of these space-times.
 In Section \ref{section-vainsh} we study how vectors contribute to the Vainshtein mechanism.
 We conclude with a section of Summary.

\section{The scalar-vector-tensor Lagrangian}\label{sec-lagrangian}

As described in the Introduction, we keep our analysis as general as possible without
focussing on any particular modified gravity model. We consider a general Lagrangian in four dimensions
for scalar, vector, and tensor DOFs that respects Galileon and Abelian gauge symmetries, and that leads
to equations of motion containing at most second derivatives on the fields. Our analysis encompasses
the phenomenology of any modified gravity model whose low energy description obeys these requirements.
Such Lagrangian contains non-linear derivative interactions involving scalar and vector DOFs, that we can
interpret as the additional gravitational degrees of freedom characterizing modifications of GR,
or alternatively as part of matter sector of the theory. The symmetries that we impose, and the physical
requirements that we demand, only  allow for a small number of terms in the effective Lagrangian ${\cal L}$
describing the non-linear interactions for our system.

The Lagrangian we will investigate has the following structure:
\be
\label{inans} {\cal L}\,=\,{\cal L}_{st}+{\cal L}_{sv}\,,
\ee
where ${\cal L}_{st}$ is the scalar tensor part, while ${\cal L}_{sv}$ is the scalar vector contribution.
Non-linear self-interactions involving tensors are described by GR and so their consequences are well known.
 For this reason we will neglect non-linear  tensor contributions in what follows: technically this condition
can be achieved by imposing that the  Lagrangian ${\cal L}$, up to total derivatives, is invariant
under linearized diffeomorphism invariance
$h_{\mu\nu}\,\to\,h_{\mu\nu}+\partial_\mu \zeta_\nu +\partial_\nu \zeta_\mu$.

In what follows we will discuss separately the two parts appearing in eq (\ref{inans}). In both cases,
the starting point is the well known fact that, in $d$ dimensions, the
following combinations of $n$ scalars is a total derivative:
\be\label{todder} {\cal L}_{der}^{(n)}\,=\,\epsilon^{\alpha_1\dots
\alpha_{d-n} \mu_1 \dots \mu_n}\,\epsilon_{\alpha_1\dots
\alpha_{d-n}}^{\hskip1cm\nu_1 \dots\nu_n}\,\,{\Pi}_{\mu_1
\nu_1}\dots{\Pi}_{\mu_n \nu_n} \ee where $\epsilon_{\mu\nu\dots} $
is the Levi-Civita antisymmetric tensor, while
$\Pi_{\mu\nu}\,=\,\nabla_\mu \partial_\nu\,\pi$, with $\pi$ a scalar
field.

The  Lagrangians in (\ref{todder}), expressed in terms of
derivatives of scalars, can be rewritten as total derivative;
on the other hand, starting from it one can obtain non-trivial field
dynamics by substituting one or more of the $\Pi_{\mu\nu}$ with other symmetric
tensors, involving for example spin-2 and spin-1 fields. This is the method we will
adopt to build the constituents of  our ${\cal L}$ in eq. (\ref{inans}).

\subsection{The scalar-tensor contribution}

The tensor scalar Lagrangian describes interactions between a tensor $h_{\mu\nu}$
(the linearized perturbation of the metric $\frac{h_{\mu\nu}}{M_{Pl}}=g_{\mu\nu}-
 \eta_{\mu\nu}$) and a scalar $\pi$. It has
the following (generalized) Galileon structure, analyzed in \cite{deRham:2010ik}:

\be {\cal L}_{st}\,=\,-\frac12 \,h^{\mu\nu}\,{\cal
E}_{\mu\nu}^{\,\,\alpha\beta}\,h_{\alpha\beta}+h^{\mu\nu}\,\sum_{n=0}^3\,
c_n\,X_{\mu\nu}^{(n)}(\Pi)\label{sctenl} \ee where ${\cal E}_{\mu
\nu}^{\alpha \beta}$ is the operator acting on $Z_{\alpha \beta}$ as
\be
 {\cal E}_{\mu \nu}^{\alpha \beta} Z_{\alpha\beta}\,=\,-\frac12\,\left(
 \Box Z_{\mu\nu}-\partial_\mu \partial_\alpha Z^\alpha_\nu-\partial_\nu \partial_\alpha Z^\alpha_\mu
 +\partial_\mu \partial_\nu Z^\alpha_\alpha-\eta_{\mu\nu} \Box Z^\beta_\beta +\eta_{\mu\nu} \partial_\alpha
 \partial_\beta Z^{\beta \alpha}
 \right).
\ee
Moreover,
\bea X_{\mu\nu}^{(n)}(\Pi)\,=\,
\epsilon_{\mu}^{\,\,\alpha_1\,...\alpha_n\,\gamma_1\,..\gamma_{3-n}}\,\epsilon_{\nu\,\,\,\;\;\;\;\;\;\;\gamma_1\,..\gamma_{3-n}}^{\,\,\beta_1\,...\beta_n}\,\Pi_{\alpha_1
\beta_1}\,...\Pi_{\alpha_n\beta_n} \eea
and $\Pi_{\mu\nu}\,=\,\nabla_\mu \partial_\nu\,\pi$, $\Pi\,=\,\Pi_{\mu}^{\,\,\mu}$.
The first term in eq. (\ref{sctenl}) is the Einstein-Hilbert (EH) Lagrangian expanded
at quadratic order in perturbations around flat space. The remaining
terms are obtained starting from the total derivative combinations
of eq. (\ref{todder}), and substituting to one of the $\Pi_{\mu\nu}$
a metric tensor $h_{\mu\nu}$.

These terms respect the Galilean symmetry
$\pi\to\pi+b_\mu x^\mu+c$,  and  lead to equations of motion with at most two  derivatives (thanks
to  the antisymmetric  properties of the Levi-Civita tensor). The Lagrangian (\ref{sctenl}) additionally  enjoys a linearized diffeomorphism invariance in the tensor sector, $h_{\mu\nu}\,\to\,h_{\mu\nu}+\partial_\mu \zeta_\nu
  +\partial_\nu \zeta_\mu$.
The resulting scalar-tensor Lagrangian contains higher order
derivative interactions for the scalar field $\pi$, that play a
crucial role for characterizing the most interesting features of
modified gravity scenarios. These interactions are controlled by
four contributions, weighted by dimensionful coefficients $c_n$
\be
c_{n}\,=\,\frac{\hat{c}_n}{\Lambda^{3(n-1)}} \,,
\ee
where
$\hat{c}_n$ is dimensionless, and $\Lambda$ some mass scale
associated with  the theory under examination.

\smallskip

Recall that the Levi-Civita tensor $\epsilon$ satisfies the following identity
\be
\epsilon_{i_1 \dots i_k~i_{k+1}\dots i_n} \epsilon^{i_1 \dots
i_k~j_{k+1}\dots j_n}
\,=\, k!~\delta^{\,\,j_{k+1} \dots j_n}_{i_{k+1} \dots i_n}
\ee
where sum over repeated indexes is assumed. The $\delta^{j_{k+1}
\dots j_n}_{i_{k+1} \dots i_n} $ denotes antisymmetrization: for
example $\delta^{ab}_{cd}=\delta^{a}_{c} \delta^{b}_d -
\delta^{a}_{d} \delta^{b}_c$. The $X^{(n)}_{\mu\nu}$ tensors
 satisfy the following recursion relation
\be\label{recrel1} X^{(n)}_{\mu\nu}\,=\,-n\,
\Pi_{\mu}^{\,\,\alpha}\,X_{\alpha \nu}^{(n-1)}+\Pi^{\alpha
\beta}\,X_{\alpha\beta}^{(n-1)} \,\eta_{\mu\nu} \ee
for $n>1$, and are symmetric and identically conserved
\be
\partial^\mu\,X^{(n)}_{\mu\nu}\,=\,0\,.
\ee

One finds for the first ones
\bea
X^{(0)}_{\mu\nu}&=&\,6\,\eta_{\mu\nu}\,,\\
X^{(1)}_{\mu\nu}&=&\,2\,\left[ \eta_{\mu\nu} \Pi- \Pi_{\mu\nu}\right]\,,\\
\eea
from which, using (\ref{recrel1}), the remaining ones can be obtained.
Notice that the contribution $h^{\mu\nu} X^{(0)}_{\mu\nu}$ corresponds to a bare cosmological constant:
we will not be interested in this and hence we will set $c_0=0$ in what follows.

It is also useful to observe that
\be
X_{\mu\nu}^{(1,\,2)}\,=\,- 2\,{\cal
E}_{\mu\nu}^{\,\,\alpha\beta}\,Z_{\alpha\beta}^{(1,\,2)}\label{diagtr}
\ee
with
\bea Z_{\mu\nu}^{(1)}\,=\,\pi\,\eta_{\mu\nu}\label{tri1}\,,
\\
Z_{\mu\nu}^{(2)}\,=\,\partial_\mu\pi \partial_\nu \pi\label{tri2}
\eea
that can be used to de-mix the kinetic terms of scalars and
tensors, and to express scalar Galileon contributions to the
Lagrangian in their original form \cite{Nicolis:2008in}.

\subsection{The scalar-vector contribution}

Various works in the past  have been dedicated to understand the
effects of vector fields in cosmology, in particular  during
inflation to build models for primordial magnetogenesis (see e.g.  \cite{Durrer:2013pga}
for a recent review). Also, non-minimal couplings between vectors and curvature can provide
models for dark energy  -- see for example \cite{Koivisto:2008xf} --  that are however often plagued by
instabilities \cite{Himmetoglu:2008zp}. It has also been shown that Horndeski vector-tensor theory
\cite{Horndeski:1976gi} leads to instabilities when applied to cosmology \cite{EspositoFarese:2009aj},
although stable regimes can be found \cite{Barrow:2012ay}.

In this work, we would like analyze models that couple vectors with scalars in a way
that preserve both Galileon and gauge symmetries. Both these symmetries might be useful to render the structure
of the Lagrangian stable under quantum corrections. In order to build the scalar-vector contribution,
we use the construction of $p$-form Galileons
\cite{Deffayet:2010zh}, and the results of \cite{Mirbabayi:2011aa}.
 The subject of couplings Galileons to gauge fields has also been investigated in \cite{Zhou:2011ix}. 
Starting from eq. (\ref{todder}), we substitute to it
one (or more) $\Pi$'s with  one (or more) symmetric tensors built up
with the vectors
\be
\label{defS} S_{\mu\nu}\,=\, \nabla_\mu A_\nu +   \nabla_\nu
A_\mu\,=\,2 \nabla_\mu A_\nu-F_{\mu\nu}\,.
\ee
In this way, one obtains  a non-trivial Lagrangian that brings dynamics to the vector
field $A_\mu$, and is characterized by equations  of motion containing at most two time derivatives
(due to the properties of the antisymmetric Levi-Civita tensor). Moreover, it is not difficult
to prove that it respects a gauge symmetry $A_\mu\,\to\,A_\mu+\partial_\mu \,\xi$ (up to total derivatives),
and the Galileon symmetry in the scalar sector.

\smallskip

It is simple to check that substituting an odd number of $S_{\mu\nu}$'s tensors in the place
of $\Pi_{\mu\nu}$'s  in eq. (\ref{todder}) provides at most total derivative
contributions, due to the properties of the Levi-Civita tensor. Substituting an even number
of $S_{\mu\nu}$ one obtains a non-vanishing result, that gives dynamics both to vector and scalar DOFs.
We focus on the four dimensional  case, in which {\it two} of the $\Pi$-tensors in eq (\ref{todder})
are substituted by the symmetric tensors $S_{\mu\nu}$.
(We do not consider the additional case in which  contractions of four $S_{\mu\nu}$ are involved,
since we checked it does not qualitatively change the results we will discuss in what follows.)
The general ghost-free vector Lagrangian that we consider, coupling scalars with vectors,  is composed
by three independent contributions:
\be\label{invel}
{\cal
L}_{sv}\,=\,S^{\mu\nu}\,\sum_{n=1}^3\,e_n\,Z_{\mu\nu}^{(n)}(S_{\rho\sigma},\Pi)\,,
\ee
with
\be
Z_{\mu\nu}^{(n)}\,=\,\epsilon_{\mu}^{\,\,\alpha_1\,...\alpha_n\,\gamma_1\,..\gamma_{3-n}}\,\epsilon_{\nu\,\,\,\;\;\;\;\;\;\;\gamma_1\,..\gamma_{3-n}}^{\,\,\beta_1\,...\beta_n}\,
S_{\alpha_1 \beta_1}\, \Pi_{\alpha_2
\beta_2}\dots\Pi_{\alpha_n\beta_n}\,.
\ee
The expressions for the $Z^{(n)}$ can be made more explicit, and read
\bea
S^{\mu\nu}\,Z_{\mu\nu}^{(1)}&=&-2 \left(S_{\mu\nu} S^{\mu\nu}-S_\mu^{\,\,\mu} S_\nu^{\,\,\nu} \right)\,,\\
S^{\mu\nu}\,Z_{\mu\nu}^{(2)}&=&-2\Pi_{\mu \nu}\,\left(S^{\mu\nu}
\,S^{\rho}_{\,\rho}- S^{\mu}_{\,\rho} \,S^{\nu\rho} \right)-
\Pi^{\rho}_{\,\rho}\,\left( S^{\mu\nu} \,S_{\mu \nu}-S^{\mu}_{\,\mu}
\, S^{\nu}_{\,\nu} \right)\,,
 \\
S^{\mu\nu}\,Z_{\mu\nu}^{(3)}&=& -2 \Pi^{\mu\nu}  \Pi^{\rho \sigma}
S_{\mu \rho} S_{\nu \sigma}+ 2 \Pi^{\mu\nu}  \Pi^{\rho \sigma}
S_{\mu \nu} S_{\rho \sigma} -4  \Pi_\mu^{\rho}  \Pi^{\mu \nu}
S_{\nu}^{\sigma} S_{\rho \sigma}+ 4  \Pi_\mu^{\mu}  \Pi^{\nu \rho}
S_{\nu}^{\sigma} S_{\rho \sigma}
+4   \Pi_\mu^{\rho}  \Pi^{\mu \nu} S_{\nu\rho} S^{\sigma}_{ \sigma}\nonumber\,\\
&&-4  \Pi_\mu^{\mu}  \Pi^{\nu \rho } S_{\nu\rho} S^{\sigma}_{
\sigma} + \left(\Pi_\mu^{\mu}\right)^2\left(S_\nu^{\nu}\right)^2-
\Pi_{\mu \rho} \Pi^{\mu\rho}\left(S_\nu^{\nu}\right)^2 - S_{\mu
\rho} S^{\mu\rho}\left(\Pi_\nu^{\nu}\right)^2+S_{\mu \rho}
S^{\mu\rho} \Pi_{\nu \sigma} \Pi^{\nu \sigma}\,.
\eea
The resulting scalar-vector Lagrangian contains higher order derivative
interactions between the vector and the scalar DOFs, that will play interesting roles in
strong coupling regimes that we will examine in the next sections. The parameters $e_n$ appearing
in the Lagrangian (\ref{invel}) are dimensionful, and can be expressed as
\be
 e_n\,=\,
 \hat{e}_n/\Lambda^{3(n-1)}
 \,,
\ee
with dimensionless $ \hat{e}_n$ and with  $\Lambda$ corresponding to some mass scale of
the theory under consideration.

\smallskip

As explained  in the appendix of \cite{Mirbabayi:2011aa}, by using
the definition (\ref{defS}), the previous Lagrangian is equivalent to a Lagrangian  in which the $S$'s
are substituted by  $F$'s, up to total derivative terms. For example, for what respect the structure of
the first contribution $e_1\,S^{\mu\nu}\,Z_{\mu\nu}^{(1)}$ to the vector Lagrangian, one finds
\bea
e_1\,S^{\mu\nu}\,Z_{\mu\nu}^{(1)}&=&e_1\,S^{\mu\nu}\,\epsilon_{\mu}^{\,\,\alpha_1\,\gamma_1\,\gamma_{2}}\,\epsilon_{\nu\,\,\,\;\gamma_1\,\gamma_{2}}^{\,\,\beta_1}\,
S_{\alpha_1 \beta_1}\,=\,-2\,e_1\,\left[S^{\mu\nu}\,S_{\mu\nu} -\left(S_\mu^\mu\right)^2\right]\,\nonumber\\&\Rightarrow&\,-2\,e_1\,F^{\mu\nu}\,F_{\mu\nu}
\eea
where to reach the second line we made integrations by parts,
neglected total derivative terms, and used the fact that
$F_{\mu}^{\mu}\,=\,0$. Hence the standard kinetic terms for the
vector are associated with the coupling
$S^{\mu\nu}\,Z_{\mu\nu}^{(1)}$ in the Lagrangian.
The remaining terms  associated with $S^{\mu\nu}\,Z_{\mu\nu}^{(2)}$
and $S^{\mu\nu}\,Z_{\mu\nu}^{(3)}$  can be also rewritten in terms
of $F_{\mu\nu}$, and lead to the combinations analyzed in
\cite{Tasinato:2012ze}:
\bea
S^{\mu\nu}\,Z_{\mu\nu}^{(2)}&\Rightarrow&\tr \Pi\,\tr F^2 -2 \,\tr
\Pi F^2\,,
\\
S^{\mu\nu}\,Z_{\mu\nu}^{(3)}&\Rightarrow& -\tr F^2\,\left[\tr \Pi^2
-\left( \tr \Pi\right)^2\right]-4 \tr \Pi\, \tr \Pi F^2 + 4 \tr
\Pi^2 F^2 +2  \tr \Pi F \Pi F\,,
\eea
where we used a more synthetic notation in terms of traces $\tr{M  N}\equiv
M_\mu^{\,\nu}\,N_{\nu}^{\,\mu}$. Hence, one can work with a vector
Lagrangian using $S_{\mu\nu}$ or $F_{\mu\nu}$ depending
on her own convenience.

\section{Maximally symmetric configurations in the vacuum and self-acceleration}\label{section-max}

After defining the Lagrangian that we will be working with, we start
to study its consequences for cosmology. In this section, we would
like to determine maximally symmetric solutions for this theory in
the vacuum. Self-accelerating solutions are included in this class,
and correspond to pure de Sitter configurations in the vacuum. Our
aim is to obtain some general lessons on the role of vector fields
for determining self-accelerating solutions in theories with
Galileon symmetry, and characterize their stability.

\smallskip

An observer, embedded into a maximally symmetric space-time, locally experiences the following
form of the metric  (for $|\vec{x}|\ll H^{-1}$, with $|\vec{x}|$ the distance from his/her position)
\be
d{s}^2\,=\,\left( 1-\frac{H^2}{2}\,x^2\right)\, \eta_{\mu\nu}\,d
x^\mu d x^\nu\label{maxsmetr}
\ee
where $H$ is the (constant) Hubble parameter: the space-time is locally  (A)dS space for
(negative) positive $H^2$. The approximation of working with a form of the metric valid only in
the region near the  observer is sufficient for our purposes.
We would like to switch on a non-trivial profile for scalar, vector,
and tensor modes so to obtain a background metric configuration as
the above. In order to do so, we consider background profiles for
the available fields as follows:
\bea
h_{\mu\nu}&=&\hat{h}_{\mu\nu}-\frac{H^2}{2}\,x^2\,\eta_{\mu\nu}\label{tenshift}\,,
\\
\pi&=&\hat{\pi}+\frac{1}{2}\,q_0\,x^2\label{scshift}\,,\\
A_\mu&=&\hat{A}_\mu+\frac{n_\mu}{2}\,x^2\label{veshift} \,,
\eea
where $H$, $q_0$ are constant numbers, while $n_\mu$ a constant
vector (that can be space-like, time-like or light-like depending on
the sign of $n^2$). The hat quantities can be interpreted as fluctuations
around the background profiles after the appropriate shifts of the fields.
In this section, we determine the conditions on the parameters $H$, $q_0$,
and $n_\mu$ to obtain a solution of the equations of motion, and the features of
the dynamics  of fluctuations around it. The field profiles
(\ref{scshift}) and (\ref{veshift}) depend on the quadratic
combination $x^2$, and are designed in such a way to generate in a
simple way the maximally symmetric solutions with metric as in eq.
(\ref{maxsmetr}). As we will see, with these choices of profiles the tadpole conditions
for the effective Lagrangian for fluctuations, which determine the background
solutions, will be satisfied by imposing simple algebraic conditions on the parameters
of the theory.

\smallskip

We substitute the shifted configurations (\ref{tenshift})-(\ref{veshift})
into our Lagrangian: these background configurations are
solutions of the background field equations if the tadpole terms in
the Lagrangian for the hat fluctuations vanish.

\subsection{Performing the shifts and obtaining
the Lagrangian for fluctuations}
In this section, we perform the shifts of eqs
(\ref{tenshift})-(\ref{veshift}), and determine the effective
Lagrangian for tensor, scalar and vector fluctuations
$\hat{h}_{\mu\nu}$, $\hat{\pi}$, $\hat{A}_\mu$ around the maximally
symmetric background of eq. (\ref{maxsmetr}).

\subsubsection*{The $\pi$ shift}

We start considering the following shifted expression for the scalar $\pi$:
\be
\pi\,=\,\hat{\pi}+\frac{1}{2}\,q_0\,x^2
\ee with
$x^2\,=\,\eta_{\mu\nu}\,x^\mu x^\nu$, and $q_0$ a constant.
Hence, the $X^{(n)}$ tensors can be expressed as
\be
X^{(n)}_{\mu\nu}\,=\,\sum_{m=0}^n\,\binom{n}{m}\,q_0^m\,
\hat{X}^{(n-m)}_{\mu\nu}\,,
\ee
where
$\hat{X}^{(n)}_{\mu\nu}\,=X^{(n)}_{\mu\nu}(\hat{\pi})$. After doing
the shift the scalar-tensor part of the Lagrangian reads
\bea
{\cal L}_{st}&=&-\frac12 \,h^{\mu\nu}\,{\cal
E}_{\mu\nu}^{\,\,\alpha\beta}\,h_{\alpha\beta}+h^{\mu\nu}\,\sum_{n=1}^3\,
c_n\, \sum_{m=0}^n\,\binom{n}{m}\,
\hat{X}^{(n-m)}_{\mu\nu}\,q_0^m\,.
\eea

\smallskip

On the other hand, a straightforward  calculation shows that the vector
Lagrangian becomes
\bea\label{newvl} {\cal L}_{sv}&=&S^{\mu\nu}\,\left( \tilde{e}_1
\,Z_{\mu\nu}^{(1)}+ \tilde{e}_2\,Z_{\mu\nu}^{(2)} +
\tilde{e}_3\,Z_{\mu\nu}^{(3)}\right) \,,
\eea
where $\pi \rightarrow \hat\pi$ in $Z_{\mu\nu}^{(n)}$ and with the tilde couplings reading
\bea
\tilde{e}_1&=&e_1+q_0 e_2+q_0^2 e_3\,,\\
\tilde{e}_2&=& e_2+2 q_0 e_3\,,\\
\tilde{e}_3&=&e_3\,.
\eea

\subsubsection*{The $h$-shift}
We consider now the shift in the tensor degree of freedom; we express
the metric $h$ tensor as
\be
h_{\mu\nu}\,=\,\hat{h}_{\mu\nu}-\frac{H^2}{2}\,x^2\,\eta_{\mu\nu}\,.
\ee
The Lagrangian expressed in terms of the variable $\hat{h}$ is
\bea
{\cal L}_{st}&=&-\frac12 \,\hat{h}^{\mu\nu}\,{\cal
E}_{\mu\nu}^{\,\,\alpha\beta}\,\hat{h}_{\alpha\beta}+3 H^2\,\hat{h}+
\left(\hat{h}^{\mu\nu}-\frac{H^2}{2} \,x^2 \,\eta^{\mu\nu}\right)
\sum_{n=1}^3\, c_n\, \sum_{m=0}^n\,\binom{n}{m}\,
\hat{X}^{(n-m)}_{\mu\nu}\,q_0^m
\\ \nonumber
&=&-\frac12 \,\hat{h}^{\mu\nu}\,{\cal
E}_{\mu\nu}^{\,\,\alpha\beta}\,\hat{h}_{\alpha\beta}+3 H^2\,\hat{h}+
\hat{h}^{\mu\nu} \sum_{n=1}^3\, c_n\,
\sum_{m=0}^n\,\binom{n}{m}\,\hat{X}^{(n-m)}_{\mu\nu}\,q_0^m -
H^2\,\hat{\pi}\,\sum_{n=1}^3\, c_n\,
\sum_{m=0}^n\,\binom{n}{m}\,\eta^{\mu\nu}
\hat{X}^{(n-m-1)}_{\mu\nu}\,q_0^m
\label{lagshif}\,.
\eea
The sums involving the binomial coefficients appearing in the
previous expression can be expanded and we obtain
\bea {\cal L}_{st}&=& -\frac12 \,\hat{h}^{\mu\nu}\,{\cal
E}_{\mu\nu}^{\,\,\alpha\beta}\,\hat{h}_{\alpha\beta}+3
H^2\,\hat{h}\nonumber
\\&&+
\hat{h}^{\mu\nu}\left[ \hat{X}^{(0)}_{\mu\nu}\,\left(
c_1 q_0+c_2 q_0^2+c_3 q_0^3 \right) +
\hat{X}^{(1)}_{\mu\nu}\,\left(c_1 +2 c_2 q_0 +3 c_3 q_0^2 \right) +
\hat{X}^{(2)}_{\mu\nu}\,\left(c_2  +3 c_3 q_0 \right) +
c_3\,\hat{X}^{(3)}_{\mu\nu} \right]\nonumber
\\&&-H^2\,\hat{\pi}\,\eta^{\mu\nu}\,
\left[ \hat{X}^{(0)}_{\mu\nu}\, \left(c_1 +2 c_2 q_0 +3 c_3 q_0^2
\right) + \hat{X}^{(1)}_{\mu\nu}\,\left(c_2  +3 c_3 q_0 \right) +
c_3 \hat{X}^{(2)}_{\mu\nu}\, \right]\,.
\eea
The scalar-vector part of the Lagrangian is not affected, since the vector does
not directly couple to the tensor.

\subsubsection*{The $A_\mu$-shift}
As a last step, we consider the vector shift (\ref{veshift}), that reads
\be\label{veshift2}
A_\mu\,=\,\frac{n_\mu}{2}\,x^2+\hat{A}_\mu
\ee
hence
\be
S_{\mu\nu}\,=\,
 \left( n_\mu \,x_\nu+n_\nu \,x_\mu\right)
+\hat{S}_{\mu\nu}
\ee
where $\hat{S}_{\mu\nu}=2\nabla_\mu \hat{A}_\nu-\hat{F}_{\mu\nu}$.
This shift only changes the structure of the scalar-vector part of the Lagrangian.
We analyze separately  the three different contributions appearing in eq.
(\ref{newvl}):

\smallskip
\noindent
 $\bullet$ $S^{\mu\nu} \,Z_{\mu\nu}^{(1)}$. This leads (up to constant terms)
 \bea
 S^{\mu\nu} \,Z_{\mu\nu}^{(1)}&=&\hat{S}^{\mu\nu} \,\hat{Z}_{\mu\nu}^{(1)}+24\,n_\mu\,\hat{A}^\mu\,.
 \eea
Notice that this piece generates a tadpole for $A^\mu$ depending on the vector $n_\mu$.

\smallskip
\noindent
$\bullet$ $S^{\mu\nu} \,Z_{\mu\nu}^{(2)}$.
This gives
\bea
 S^{\mu\nu} \,Z_{\mu\nu}^{(2)}&=&\hat{S}^{\mu\nu} \,\hat{Z}_{\mu\nu}^{(2)}+ 4\left(
 \,n_\mu\,\hat{A}^\mu\,\Box \hat{\pi}-
 n_\mu \hat{A}_\rho\,\hat{\Pi}^{\rho \mu}
 \right)+12\, \hat{\pi} \,n_\mu n^\mu\,.
\eea
This piece generates a tadpole for $\pi$, and a quadratic coupling
between scalar and vector.

\smallskip
\noindent
 $\bullet$ $S^{\mu\nu} \,Z_{\mu\nu}^{(3)}$. This gives
\bea
 S^{\mu\nu} \,Z_{\mu\nu}^{(3)}&=&\hat{S}^{\mu\nu} \,\hat{Z}_{\mu\nu}^{(3)}- 2\,\left(n^\mu\,\hat{A}^\nu
 +n^\nu\,\hat{A}^\mu
  \right)\,
  \hat{X}^{(2)}_{\mu\nu}
  +4
n^2 \hat{\pi} \Box \pi+4\left( n^\mu \partial_\mu \hat{\pi}\right)^2
 \,.
 \eea
 \bigskip
The complete vector Lagrangian can then be obtained by plugging these
different pieces into eq. (\ref{newvl}). One finds
\bea
\label{veclag}
{\cal L}_{sv}&=& \tilde{e}_1\left(\hat{S}^{\mu\nu} \,\hat{Z}_{\mu\nu}^{(1)}+24\,n_\mu\,\hat{A}^\mu \right)\nonumber\\
&&+\tilde{e}_2 \left( \hat{S}^{\mu\nu} \,\hat{Z}_{\mu\nu}^{(2)}+
4\left(
 \,n_\mu\,\hat{A}^\mu\,\Box \hat{\pi}-
 n_\mu \hat{A}_\rho\,\hat{\Pi}^{\rho \mu}
 \right)+12\, \hat{\pi} \,n_\mu n^\mu
\right)\nonumber\\
&&+\tilde{e}_3 \left( \hat{S}^{\mu\nu} \,\hat{Z}_{\mu\nu}^{(3)}-
2\,\left(n^\mu\,\hat{A}^\nu
 +n^\nu\,\hat{A}^\mu
  \right)\,
  \hat{X}^{(2)}_{\mu\nu}
  +4
n^2 \hat{\pi} \Box \hat{\pi}+4\left( n^\mu \partial_\mu
\hat{\pi}\right)^2  \right)
\eea
Notice that the Lagrangian for fluctuations explicitly depends on the direction $n^\mu$ along which we turn on the vector profile 
(see eq. (\ref{veshift2})). Hence while the  background configurations  for the metric and the scalar 
 are isotropic, the dynamics
of scalar and vector fluctuations depend 
on the particular direction along the vector background.

\subsubsection*{Imposing the tadpole conditions}

A necessary and sufficient condition to determine background solutions is to cancel the tadpole terms
depending on tensor, scalar, and vector hat fluctuations. These read
\bea &&\text{(tensor)}\hskip0.9cm
 H^2+2\,\sum_{n=1}^3\,
c_n\,q_0^n\,=\,0\hskip1.6cm\Rightarrow\hskip0.4 cm -\frac{H^2}{2}\,=\,c_1 q_0+c_2 q^2_0+c_3 q^3_0\label{condi1}\,,\\
&&\text{(scalar)}\hskip0.9cm
 H^2\,\sum_{n=1}^3\,
c_n\, n\,q_0^{n-1}+12\,e_2\, n^2
\,=\,0\hskip0.1cm\Rightarrow\hskip0.5cm
H^2\left(c_1+2\,q_0\,c_2+3\,q_0^2\,c_3\right)\,=\,-{12} \,e_2\,n^2
\label{condi2}\,,
\\&&
\text{(vector)}\hskip0.9cm \left( e_1+q_0 e_2+q_0^2 e_3
\right)\,n_\mu
\hat{A}^{\mu}\,=\,0\label{condi3}\hskip0.5cm\Rightarrow\hskip0.5cm
\tilde{e}_1\,=\, \left( e_1+q_0 e_2+q_0^2 e_3 \right)\,=\,0
\hskip0.3cm {\text{or}}
 \hskip0.3cm n_\mu\,=\,0  \,.
\eea

\bigskip

As anticipated above, these are algebraic equations between the quantities appearing
in the Lagrangian for fluctuations. Choosing parameters such to satisfy these conditions,
we find maximally symmetric space-times in the vacuum, around which  the dynamics
of perturbations preserve Galilean and gauge symmetries.
The three conditions above fix the numerical quantities $q_0$, $H_0$ and $n^2$ completely.
 We can find different branches of solutions of the previous system
of equations, that we will discuss in what follows. 
As we will see,  although we can switch on a vector profile along a direction $n^\mu$, nevertheless we
 will be able to find isotropic (and maximally symmetric) solutions for the metric. This is due to the particular
 derivative couplings of the scalar to the vector,  that can allow  us to solve  eq. (\ref{condi3}) with $n_{\mu}\neq 0$.

\subsection{The maximally symmetric background solutions}\label{sec-msbsol}
 The different branches of solutions to the set of equations (\ref{condi1})-(\ref{condi3}) are:

\smallskip

\noindent
{\bf 1. First branch}:

\noindent
Suppose that equation (\ref{condi3}) admits real solutions for $q_0$
by imposing $\tilde{e}_1\,=\,0$. Then, it determines up to two real solutions for $q_0$. Plugging one of these
solutions in (\ref{condi1}) we determine $H^2$. Plugging these results in  (\ref{condi2})  one finally determines $n^2$. Notice that $H^2$ and $n^2$ can  have either sign: if $H^2\,>\,0$ one obtains de Sitter space and a self-accelerating configuration in the vacuum. This branch is characterized by a background vector field turned on  with $n_\mu\neq 0$,
that however does not break the isotropy of three dimensional space-time, nor breaks the vector gauge symmetry.
Let us emphasize that this branch is characterized by the condition $\tilde{e}_1\,=\,0$: this implies the vanishing of
the standard kinetic term for the vector fluctuations $\hat{A}_\mu$, that is  proportional to $S^{\mu\nu} \hat{Z}_{\mu\nu}^{(1)}$ in the Lagrangian (\ref{veclag}). On the other hand, the vector fluctuations can acquire dynamics through coupling with the scalar fluctuations $\hat{\pi}$, as we will see below.

\smallskip
\noindent
{\bf 2. Second branch}:
\noindent We can also recover well-known scalar Galileon maximally symmetric
solutions with vector field turned off. Choose $n^\mu=0$, hence  (\ref{condi3}) does
not give any constraint on $q_0$. If  (\ref{condi2})  admits at least one  real solution for $q_0$,
we can use its value in (\ref{condi1}) to determine $H^2$.  If $H^2>0$, we obtain
the self-acceleration.  In this case the kinetic terms for the vector fluctuations are generally
 not
vanishing. 

\smallskip
\noindent
{\bf 3. Third branch}:
\noindent
The last option is to turn off the gauge field, $n^\mu=0$, and choose the Minkowski space with $H=0$.
Hence (\ref{condi2})  and  (\ref{condi3}) are  automatically  satisfied. Then (\ref{condi1}),
when admitting  real solutions, fixes $q_0$.

\bigskip

The first branch of solutions is new, and specific to the case of having a vector field turned on (although 
 similar configurations have been already studied in massive gravity \cite{Koyama:2011wx}).
The other two branches were already known in the literature, at least for the specific set-up of massive
gravity \cite{deRham:2010tw} (while the case with $c_3 = 0$ had been already investigated in \cite{Nicolis:2008in}).
Notice that we can have intermediate situations in which different branches are connected.
Suppose that condition (\ref{condi3}) is satisfied with $\tilde{e}_1\,=\,
\left( e_1+q_0 e_2+q_0^2 e_3 \right)\,=\,0$, and the value of $q_0$
satisfying this condition also satisfies  (\ref{condi2})  with
$n^2=0$ (but with the (null-like) vector $n_\mu$ not necessarily
vanishing). This configuration continuously connects the first and
second branches.  A similar situation can be realized, for example,
in massive gravity \cite{deRham:2010tw,Koyama:2011wx,Tasinato:2012ze}.

\subsection{An instability   around the first branch of maximally symmetric solutions}

Let us focus on the first branch with the vector fields turned on to study the
dynamics of the fluctuations. After imposing the conditions to remove the tadpoles, the
quadratic contributions to the complete Lagrangian result

\bea
{\cal L}_{quadr}&=&-\frac12 \,\hat{h}^{\mu\nu}\,{\cal
E}_{\mu\nu}^{\,\,\alpha\beta}\,\hat{h}_{\alpha\beta}
-\frac{12\,n^2}{H^2}
\,\hat{h}^{\mu\nu}\hat{X}_{\mu\nu}^{(1)}
-6 H^2 \left( c_2+3 c_3 q_0\right)\,\hat\pi \Box \hat\pi\nonumber\\
&&+4\,\tilde{e}_3\,\left( n^2 \hat\pi \Box \hat\pi+\left( n^\mu
\partial_\mu \hat\pi\right)^2 \right)
+4 \tilde{e}_2 \,\left(
 \,n_\mu\,\hat{A}^\mu\,\Box \hat\pi-
 n_\mu \hat{A}_\rho\,\hat\Pi^{\rho \mu}
 \right)\,.\label{qua1}
\eea
Let us emphasize again that in this branch the vector tadpole cancelation,
associated with condition (\ref{condi3}), implies that the vector field has no standard kinetic term.
On the other hand, the vector acquires  a coupling with the scalar at quadratic order in perturbations
(if $\tilde{e}_2$ is non vanishing, as we will suppose from now on) that depends  on the background
vector profile $n_\mu$. Notice that the previous quadratic contribution to the Lagrangian is
linear on $\hat{A}_\mu$. On the other hand, higher order contributions to the Lagrangian will also
include terms quadratic in the vector field.

The quadratic Lagrangian for tensor and scalar can then be
diagonalized with the standard field transformation of
$\hat{h}_{\mu\nu}$ to $ \tilde{h}_{\mu\nu}$
\be
\hat{h}_{\mu\nu}\,=\, \tilde{h}_{\mu\nu}
+\frac{6\,n^2}{H^2}
\,\hat\pi\,\,\eta_{\mu\nu}
\ee
finding
\bea {\cal L}_{quadr}&=&-\frac12 \,\tilde{h}^{\mu\nu}\,{\cal
E}_{\mu\nu}^{\,\,\alpha\beta}\,\tilde{h}_{\alpha\beta} + \left[
\frac{108\,(n^2)^2}{H^4}
%
-6 H^2 \left( c_2+3 c_3 q_0\right) \right]\,\hat\pi \Box \hat\pi
\nonumber\\
&& -2\tilde{e}_3\,\left( n_\mu \partial_\nu \hat\pi -n_\nu
\partial_\mu \hat\pi  -\frac{\tilde{e}_2}{2 \tilde{e}_3}
\,\hat{F}_{\mu\nu}\right)^2
+\frac{\tilde{e}_2^2}{2
\,\tilde{e}_3}\,\hat{F}_{\mu\nu}^2\label{qua2} \,.
\eea
The scalar-vector coupling at quadratic order (associated with the first
parenthesis in the second line of the previous formula (\ref{qua2}))
cannot be removed by a simple local field redefinition. But the
structure of the Lagrangian is sufficiently simple to exhibit an
instability.

\bigskip

For simplicity, let us focus on purely time dependent perturbations,
with $\hat{\pi}=\hat{\pi}(t)$, and $\hat{A}_\mu\,=\,(0, \, A_r(t),\,
0,\,0)$. This Ansatz for the fluctuations is very simple, but it is sufficient
for our purpose. Focus on the  scalar-vector part of Lagrangian (\ref{qua2}); it
can be rewritten as
\bea
{\cal L}_{quadr}&=&\left[ \frac{108\,(n^2)^2}{H^4}
%
-6 H^2 \left( c_2+3 c_3 q_0\right) +4
\tilde{e}_3\,n_r^2\right]\,\dot{\hat\pi}^2 +4
\tilde{e}_2\,n_r\,\dot{A}_r\,\dot{\hat\pi} \,. \label{qua3}
\eea
Calling ${\cal Q}$ the part in square parenthesis of the previous
equation, one finds
\bea
{\cal L}_{quadr}&=&{\cal Q}\,\left(\dot{\hat\pi}+{\frac{2\tilde{e}_2
n_r}{{\cal Q}}}\,\dot{A}_r \right)^2 -\frac{4\tilde{e}^2_2 n^2_r}{{\cal
Q}}\,\dot{A}_r^2 \,. \label{qua4}
\eea
Hence, when ${\tilde e}_2\neq0 $, one always finds an instability on the scalar,
or on the vector sectors (depending on the sign of ${\cal Q}$)
for the first branch of solutions~\footnote{The case ${\tilde
e}_2=0$ can be studied using the Hamiltonian approach applied in
\cite{Tasinato:2012ze,Tasinato:2013rza} for the special case of
massive gravity.}.

\smallskip

The conclusion is that the first branch of maximally symmetric
configurations for our theory, that admits a non-trivial profile for
the vector field, is generically unstable. Vector or scalar
fluctuations have the wrong sign for the kinetic terms. The same
consideration holds for set-ups  that interpolate between the first
and second branches, as the ones discussed at the end of Section
\ref{sec-msbsol}, in which a light-like vector field can be turned on.

\smallskip

The general lesson is that, when considering modified gravity scenarios that
rely on Galileon symmetries, one has to pay extra care to the dynamics of
vector fluctuations, in particular around the branches of self-accelerating
solutions in which the standard vector kinetic terms vanish. Indeed, these
configurations are generally plagued by instabilities associated with higher order
Galileon interactions between vector and scalars. This fact has been pointed out
in \cite{Tasinato:2012ze,Koyama:2011wx} for the special case of massive gravity in
decoupling limit: the results of the present paper generalize the analysis to a broader
context and provide the tools to analyze this issue in more general set-ups
respecting Galileon and gauge symmetries.

\section{Spherically symmetric configurations: how vectors contribute to the Vainshtein mechanism}\label{section-vainsh}

Vector degrees of freedom derivatively coupled to scalars can
contribute to the screening mechanisms that characterize the most
interesting modified gravity scenarios.

\smallskip

In this section, we will analyze spherically symmetric solutions
around a given source in the theory we are considering.
Usually, for simplicity in treating modified gravity scenarios one
makes the hypothesis that the only degrees of freedom coupling to the source
are the tensor, via a minimal coupling $h_{\mu\nu} T^{\mu\nu}$, and the scalar,
which couples to the trace of the energy momentum tensor as $\pi T$.
On the other hand, since vector degrees of freedom are normally contained in the low-energy spectrum
of gravitational interactions,  they can couple directly to the source in a way that respects
the symmetries of the theory. We propose to consider the case in which, besides the usual energy
momentum tensor, sources are also characterized by gravitational vector currents $J^\mu$,
associated with a gravitational vector coupling of the form $A_\mu J^\mu$.
If the vector currents are conserved, such couplings respect the Abelian gauge symmetry
$A_\mu\to A_\mu+\partial_\mu \xi$ of the Lagrangian.
As we will see later, the vector charge influences the realization of the Vainshtein effect, and
provides new interesting environmental effects that change the gravitational interactions
around a spherically symmetric source. (See also \cite{Jimenez:2012ph} for a discussion of a screening mechanism
in a theory involving vectors.)

\smallskip

We look for static spherically symmetric configurations around
Minkowski space~\footnote{The same analysis holds also around the
maximally symmetric second branch of solutions discussed in the
previous section.}. In order to study these configurations, we
re-express the scalar and vector Lagrangians in  a  more convenient
form.  We set to zero the coupling $c_3$ of the scalar Galileon
Lagrangian of eq. (\ref{sctenl}), since it has been shown in
\cite{Koyama:2013paa} that these couplings lead to instabilities around
spherically symmetric solutions. After performing a proper diagonalization
procedure using the relations (\ref{diagtr})-(\ref{tri2}), the scalar Lagrangian
can be written as
 \be
 {\cal L}_{sc}^{tot}\,=\,\sqrt{-g}\,\left[d_2\,{\cal L}_{sc}^{(2)}+d_3\,{\cal L}_{sc}^{(3)}+d_4\,{\cal L}_{sc}^{(4)}\right]
 \ee
with \bea
{\cal L}_{sc}^{(2)}&=&\frac12\,\partial \pi \cdot \partial \pi \,,\\
{\cal L}_{sc}^{(3)}&=&\frac12\, \left(\tr{ [\Pi]}\right)\,\partial \pi \cdot \partial \pi \,,\\
{\cal L}_{sc}^{(4)}&=&\frac14\,\left( \left( \tr{[\Pi]}\right)^2
 \partial \pi \cdot \partial \pi
 -2 \tr\left[ \Pi\right]\,\partial \pi \cdot \Pi\cdot \partial \pi -
  \tr\left[ \Pi^2\right]\,\partial \pi \cdot \partial \pi +\partial \pi \cdot \Pi^2\cdot \partial \pi
\right)\,, \eea
where the $d_i$ are suitable linear combinations of
the original parameters $c_i$ appearing in eq. (\ref{sctenl}). These
couplings can be expressed as
\be
d_i\,=\,\frac{\hat{d}_i}{\Lambda^{3(i-2)}}\label{dless1}\,,
\ee
where the $\hat{d}_i$ are dimensionless quantities while $\Lambda$ is a scale
of dimension of a mass, whose value depends on the theory under consideration.
$\Pi$ corresponds to the matrix $\Pi_{\mu\nu}$, and we indicate
traces with $\tr \left[ \dots \right]$. The vector Lagrangian is
expressed in terms of $F_{\mu\nu}\,=\,\nabla_\mu A_\nu-\nabla_\nu
A_\mu$ as
\bea\label{lagvec} {\cal
L}_{sv}^{tot}\,=\,\sqrt{-g}\,\left[ e_2\,{\cal L}_{sv}^{(2)}
+e_3\,{\cal L}_{sv}^{(3)}+{e_4}\,{\cal L}_{sv}^{(4)}\right] \,,
\eea
with
\bea
{\cal L}_{sv}^{(2)}&=& -\tr{[F^2]} \,,\\
{\cal L}_{sv}^{(3)}&=&-\left( \tr[\Pi]\,\tr[ F^2] -2 \,\tr[ \Pi
F^2]\right) \,,
\\
{\cal L}_{sv}^{(4)} &=&-\frac14\,\left( -\tr [F^2]\,\left[\tr[
\Pi^2] -\left( \tr [\Pi]\right)^2\right]-4 \tr [\Pi]\, \tr[ \Pi F^2]
+ 4 \tr[ \Pi^2 F^2] +2  \tr[ \Pi F \Pi F] \right) \,.
\eea

The parameters $e_n$ can be associated with linear combinations of
the parameters $e_i$ appearing in the Lagrangian (\ref{invel}) that
was expressed in terms of the tensor $S_{\mu\nu}$. These parameters
can be expressed as
\be
e_n\,=\,\frac{\hat{e}_n}{\Lambda^{3(n-2)}}\label{dless2}\,,
\ee
where the $\hat{e}_n$ are dimensionless while $\Lambda$ is a scale of
dimension of a mass.

To the previous Lagrangians we then add contributions that control
couplings of tensor, scalar, and vectors to the source. We assume
that a source is characterized by a conserved energy momentum tensor
$T_{\mu\nu}$, and a conserved  vector current $J_\mu$. The couplings
we consider are
\be {\cal
L}_{coupl}\,=\,\sqrt{-g}\,\left[\frac{1}{M_{Pl}}\,h_{\mu\nu}\,T^{\mu\nu}
+\frac{1}{2\,M_{Pl}}\,\pi\,T+A_\mu\,J^{\mu}\right]\,.
\ee

\smallskip

We parameterize flat Minkowski space in spherical coordinates: we will then
include the overall factor $\sqrt{-g}\,=\,r^2\,\sin{\theta}$ to the previous Lagrangians.
Focussing on static spherically symmetric configurations, we drop
the explicit dependence on time, and hence focus on the  component
$A_\mu\,=\,(A_0(r),0,0,0)$ for the vector, and $\pi(r)$ for the
scalar.

The scalar Lagrangian (including the coupling with source) in this
spherically symmetric case reads
\be
\frac{{\cal
L}_{sc}}{\sin{\theta}}\,=\,-\frac{d_2}{2}\,r^4\,\left(\frac{\pi'}{r}\right)^2-\frac{2\,d_3}{3}\,r^4\,\left(\frac{\pi'}{r}\right)^3
-\frac{d_4}{2}\,r^4\,\left(\frac{\pi'}{r}\right)^4+\frac{r^2}{2
M_{Pl}}\,\pi\,T \,.
\ee
We choose the energy momentum tensor of a
spherically symmetric, point-like source of mass $M$. The trace of
it is given by
\be
T
\,=\,M \,\frac{\delta(\vec{r})}{r^2}\,=\,-M
\,\frac{1}{r^2}\,\partial_r\,\left[r^2\,\partial_r\left(\frac{1}{r}\right)
\right]\,.
\ee
Then the scalar Lagrangian, upon integration by parts,
becomes
\be
\frac{{\cal
L}_{sc}}{\sin{\theta}}\,=\,-\frac{d_2}{2}\,r^4\,\left(\frac{\pi'}{r}\right)^2-\frac{2\,d_3}{3}\,r^4\,\left(\frac{\pi'}{r}\right)^3
-\frac{d_4}{2}\,r^4\,\left(\frac{\pi'}{r}\right)^4-\,\frac{\pi'}{r}\,\left(\frac{M\,r}{2\,M_{Pl}}\right)
\,.
\ee

\smallskip

As explained above, we assume that each source couples not only to
the scalar, but also to the vector field via a current in the form
$A_\mu\,J^\mu$. Our  point-like  source is characterized by a
non-vanishing  dimensionless  vector charge $Q_0$. For a spherically
symmetric, static source  the vector current reads
$J^0\,=\,2\, Q_0 \,\delta^{(3)}(x)$.

The vector Lagrangian, after integrating by parts, is given by
\be
\frac{{\cal L}_{sv}}{\sin\theta}\,=\,A_0'(r)^2\,r^2\,\left(2 e_2+ 4
e_3 \frac{\pi'}{r}+e_4 \frac{\pi'^2}{r^2}\right)-2\,Q_0\,A_0'(r)\,
\,.
\ee
We can solve the equation of motion corresponding to
$A'_0(r)$:
\be
\label{elefi} A_0'(r)\,=\,\frac{Q_0}{r^2\,\left(2 e_2+ 4 e_3
\frac{\pi'}{r}+e_4 \frac{\pi'^2}{r^2}\right)}\,.
\ee
This is proportional to the `electric' part of the vector field  strength
associated with a point-like charge $Q_0$. Since no time
derivatives are involved, we can plug the result for $A_0'(r)$ back
in the vector Lagrangian, finding
\be {\cal
L}_{sv}\,=\,-\frac{Q_0^2}{r^2\,\left(2 e_2+ 4 e_3 \frac{\pi'}{r}+e_4
\frac{\pi'^2}{r^2}\right)} \,.
\ee
When added to the scalar Galileon terms, we find the following algebraic equation
of motion for the quantity $y=\pi'/r$ that controls the scalar field,
in the presence of a source with mass and vector charge
\be
d_2 \,y+2\,d_3\, y^2+2\,d_4\,y^3-\frac{\left(4
e_3 +2 e_4 y\right)\,Q_0^2}{r^{6}\,\left( 2 e_2+4 e_3 y+e_4
y^2\right)^2}\,=\,\frac{M}{M_{Pl}\, r^3}
\label{geneomsc}\,.
\ee
We study now the system in two different cases. The case in which only
the cubic Galileon is included (setting the quartic couplings $d_4$
and $e_4$ to zero in the equation above)
will be discussed in the next section.
The case in which also
the quartic Galileon is switched on is conceptually very similar, and
its analysis is relegated to the Appendix.  For simplicity, we will not
consider here the effect of the quintic Galileon in this work.

\subsection{The cubic Galileon}
Let us start our discussion with the case of cubic Galileon: set $d_4\,=\,e_4\,=\,0$
in eq. (\ref{geneomsc}). The scalar equation is
\be\label{simpeq}
d_2 \,y+2\,d_3\, y^2-\frac{4 e_3
\,Q_0^2}{r^{6}\,\left( 2 e_2+4 e_3 y\right)^2}\,=\,\frac{M}{M_{Pl}\,
r^3}\,.
\ee
Far from the source, $r\gg1$, $y$ is small and the
solution of the previous equation is $y= M/( M_{Pl} d_2 \,r^3)$, which
implies $ \pi'= M/( M_{Pl}\,d_2 \,r^2)$ and $A_0' = Q_0/(2 \,e_2 \,
r^2)$.  Hence scalar and vectors mediates  fifth, long-range forces that
lead to a modification with respect to GR predictions. Notice that
the expression for $\pi$ does not depend on the vector charge $Q_0$
in this large $r$ limit.

\smallskip

More interesting to us is what happens in proximity of the source:
hence $y$ becomes large, and (\ref{simpeq}) admits two branches of
solutions (recall the definition of the dimensionless hat quantities eqs
(\ref{dless1}), (\ref{dless2}))
\bea
y&=&\sqrt{\frac{M}{4\,\,d_3\,M_{Pl}\,r^3}}\,\sqrt{1\pm\sqrt{1+\frac{2
d_3 M_{Pl}^2 \,Q_0^2}{e_3 M^2}}}
\\
&\equiv&\frac{\hat{d}_2\,\Lambda^3 \,\left(r_V^\pm\right)^{3/2}}{\hat{d}_3\,\,r^{3/2}}
\label{addd2} \,,
\eea
with
\be\label{chrvpm}
r_V^{\pm}\,\equiv\,\frac{1}{\Lambda}\,
\left[\frac{\hat{d}_3\,M}{4\,\hat{d}_2^2\,M_{Pl}}\,\left(
1\pm\sqrt{1+\frac{2 \hat{d}_3 M_{Pl}^2 \,Q_0^2}{\hat{e}_3 M^2}}
\right) \right]^{\frac13}\,,
\ee
where the Vainshtein radius $r_V$ is defined in such a way to correspond
to the scale at which the non-linear terms in the scalar equations
of motion become important. Indeed when $r=r_V$, one finds that the value of
$y$ is  $y=\hat{d}_2 \Lambda^3 / \hat{d}_3$. This is the scale at which the
second term in the expression (\ref{geneomsc}) becomes comparable to the first term.
This is why we included the additional factors in eq. (\ref{addd2}).

At short distances $r\ll1$ we get the following behavior for the scalar field
$\pi$ and $A_0$,
\bea
\pi(r)&=&\frac{2 \hat{d}_2}{\hat{d}_3}\,\Lambda^3\,
\left(r_V^{\pm}\right)^\frac32\,\sqrt{r} \,,
\\
A_0(r)&=&\frac{Q_0\,\hat{d}_3\,\sqrt{r}}{2\,\hat{e}_3 \,\hat{d}_2\,
\left(r_V^{\pm}\right)^\frac32} \,.
\eea
Since $\pi', A_0'\,\propto\,1/\sqrt{r}$ the scalar and vector contributions are
much weaker than the usual gravitational one in proximity of the source: the Vainshtein
mechanism is at work and GR is recovered nearby a spherically symmetric source.

As we will see explicitly in a moment, the requirement of stability
of our configuration demands that the parameters $\hat{d}_2$,
$\hat{d}_3$, $\hat{e}_2$, $\hat{e}_3$ are non-negative.  The
obvious requirement of having $r_V^\pm>0$ selects only the positive
branch in the above choice (\ref{chrvpm}) (at least for $M>0$) hence
we will focus on this case from now on. Then,
the value of the Vainshtein radius depends not
only on the mass of the object, but also on how much the object is
coupled to the vector fields. This fact could play an important role
for analyzing the effective theory of fluctuations around a source
and increase the effective cut-off for low energy theory of
perturbations, as discussed for example in  \cite{Burrage:2012ja}.
Hence higher order Galilean interactions involving vectors, as the ones we
consider, render environmental effects richer and subtler.

Before briefly discussing some phenomenological consequences of
these  findings, let us analyze more in detail the stability under
small fluctuations of the spherically symmetric backgrounds we have
determined. We focus on spherically symmetric scalar and vector perturbations,
which respect the spherically symmetric Ansatz in terms of
the criteria used in \cite{Koyama:2011wx}:
\bea
\pi&=&\bar{\pi}(r)+\varphi(t,r)\,,\\
A_{\mu}&=&\left( \bar{A}_0(r)+\delta A_0(t,r),\,\delta
A_r(t,r),\,0,\,0\right)
\,,\\
 F_{tr}&=&\delta \dot{A}_r(t,r)-\delta {A}_0'(t,r)\,.
\eea
The bars denote background quantities. We limit our attention to the spherically symmetric
case to be able to treat fully analytically the system of coupled scalar and vector equations
of motion for the fluctuations.

After a straightforward de-mixing procedure of scalar from vectors,
that involves a field redefinition
\be
 F_{tr}\,\to\,-\frac{2\,e_3\,\bar{A}_0'\,\varphi'}{e_2\,r+2\,e_3\,\bar{\pi}'}\,+\,\tilde{F}_{tr}\,,
\ee
we obtain the following Lagrangian for spherically symmetric
quadratic fluctuations around our spherically symmetric background
\be
S_{\varphi}\,=\,\frac12\,\int\,d^4 x\,\sqrt{-g}\,\left[
K_t(r)\,\dot{\varphi}^2-K_r(r)\,{\varphi'}^2-K_A(r)\,\tilde{F}_{tr}
\tilde{F}^{tr}\right]\,. \ee The functions  $K_i$ have to be
positive in order to obtain a stable configuration. They read
\bea K_t&=&\,{d_2}+4\,d_3\,\frac{\bar{\pi}'}{r}
+2\,d_3\,\bar{\pi}''\,,
\\
K_r&=&\, {d_2}+4\,d_3\,\frac{\bar{\pi}'}{r} +\frac{4\,e_3^2
\,Q_0^2}{r^6\left(e_2+2 e_3 \bar{\pi}'/r \right)^3} \,,
\\
K_A&=&  e_2+2\,e_3\,\frac{\bar{\pi}'}{r}\,.
\eea
Nearby the source, the quantity $\bar{\pi}'/r$ is large: in order to have positive
$K_i$, we demand that $d_3$, $e_3$ are positive. Far from the
source, instead, $\bar{\pi}'/r$ is small: we have to demand that
$d_2$, $e_2$ are positive. This implies that a vector charge
increases the size of the function $K_r$ above nearby the source,
while it gives only negligible contributions far from it. Notice that these results,
 as anticipated,  require that we can only take the positive sign in
the option  for the Vainshtein radius 
in eq. (\ref{chrvpm}). 

Computing the speed of radial scalar fluctuations in proximity of
the source we find
\be c_r^2\,\equiv\,\frac{K_r}{K_t}\,=\,
\frac43\,\left( 1+\frac{{\cal Y}}{\left( 1+\sqrt{1+{\cal
Y}}\right)^2}\right)
\ee
with
\be {\cal Y}\,\equiv\,\frac{2
\hat{d}_3 M_{Pl}^2\,Q_0^2}{M^2\,\hat{e}_3}
\ee
Hence, being ${\cal Y}\,\ge\,0$, the already superluminal speed of radial fluctuations
is further increased by the presence of the vector charge.
Far from the source, the vector charge gives negligible contributions and one recovers the well-known
predictions of standard scalar Galileon models. We conclude that the vector charge does not help
to solve the issue of superluminal propagation in cubic Galileon theories.

\subsection{Phenomenological considerations}
Let us make some simple phenomenological considerations on the results obtained so far.
Under the hypothesis that the vector fields are part of the spectrum of gravitational DOFs,
a non-vanishing charge $Q_0$ would be  associated with  gravitational  vector interactions.
The vector charge modifies the expression for the Vainshtein radius for the scalar interaction,
which is bounded from below no matter how small the mass of the source is:
\be\label{rvlimit}
r_V^3\,\ge\,
\frac{1}{\Lambda^3}\,
\frac{\hat{d}_3^{3/2}\,|Q_0|}{\sqrt{8}\,\hat{d}_2^2\,\hat{e}_3^{1/2}}
\ee
More in general, if the dimensionless quantity
\be
{\cal Y}\,\equiv\,\frac{2 \hat{d}_3 M_{Pl}^2\,Q_0^2}{\hat{e}_3\,M^2}\,,
\ee
is much larger than one, then the expression for the Vainshtein
radius is sensitive to the vector charge $Q_0$, and saturates the
inequality (\ref{rvlimit}) in the limit ${\cal Y}\to\infty$.
Notice that ${\cal Y}$ does not depend on the scale $\Lambda$, and is
proportional to the (typically very large) ratio $M_{Pl}^2/M^2$.

The ratio of the Vainshtein radii, calculated respectively in the
limits of large and small ${\cal Y}$, reads
\be
\frac{r_V^{{\cal Y}\to\infty}}{r_V^{{\cal Y}\to0}}\,=\,{{\cal Y}^\frac16}\,.
\ee
Hence the Vainshtein radius can increase considerably in the
presence of a vector charge, potentially changing
the predictions of modified gravity scenarios based on the Vainsthein mechanism.


\bigskip

Until now, our considerations were made under the hypothesis that
the vectors under consideration are part of the gravitational sector
of the theory. In this case, we found that vector gravitational
interactions, as well as scalar interactions, are screened well
inside the Vainshtein  radius, and GR predictions are well
recovered nearby a source.

For the remaining part of this section we would like to consider the
different, alternative perspective that the vectors we are analyzing are part of the matter sector,
and not of the gravitational sector of the theory. We consider a purely scalar Galileon theory describing
a theory of  modified gravity, that couples with standard electromagnetism.
Higher order interactions between the scalar $\pi$ and electromagnetism, with a structure
described by Lagrangian (\ref{lagvec}), are allowed by the symmetries of the theory: the coefficient
$\hat{e}_3$ and the scale $\Lambda$ should then be constrained in such a way to
agree with the very accurate experimental tests of electromagnetic interactions and quantum
electrodynamics. This is an interesting topic, that however we will not analyze in this work.
Here, we would like only to point out how the electric charge of a body changes
the size of its Vainsthein radius, using the results that we obtained above.

Consider for definiteness an electron: it has an electromagnetic vector charge, as well as a mass.
In appropriate units, the ratio between electron charge $q_e$ and electron mass $m_e$ (including a Planck
mass to render this quantity  dimensionless) is
\be\label{ratioele}
\frac{ M_{Pl}^2\,q_e^2}{m_e^2} \simeq 10^{42}
\ee
and this huge number reflects the well known fact that the
relative strength of  the electromagnetic force is much larger than
the one of gravitational  interaction. The ratio (\ref{ratioele})
enters in the expression for the Vainsthein radius for a charged
body. If the parameters $\hat{e}_i$, $\hat{d}_i$ are not exceedingly small, for the
considerations we made above this implies that the size of Vainshtein radius for an electron
is independent of the electron mass, and reads
\be
r_V\,=\,\frac{1}{\Lambda}\, \,\left[
\frac{\hat{d}_3^{3/2}\,q_e}{\sqrt{8}\,\hat{d}_2^2\,\hat{e}_3^{1/2}}\right]^\frac13
 \simeq\,10^{10}\,
 \left[  \frac{\hat{d}_3^{3/2}}{\sqrt{8}\,\hat{d}_2^2\,\hat{e}_3^{1/2}}\right]^\frac13
 \,
\left(\frac{m_e}{\Lambda}\right)\,\,
r_e
\,.
\ee
where $r_e\,=\,10^{-15}$ meters is the scale of the classical radius of the electron. Hence
we learn that for scales $\Lambda$  of order of the electron mass and choosing the
couplings not exceedingly small, the Vainshtein radius
would be much larger than the classical electron radius.

On the other hand, we have to take into account that also the
electromagnetic force is changed when coupling the scalar to the
electromagnetic field.
 Recall that   the spherically symmetric electric
field produced by the electron, for the case
of cubic Galileon we are focussing on in this section,
  is given by the expression
(\ref{elefi}) that
reads
\be
A_0'(r)\,=\,\frac{Q_0}{r^2\,\left(2 e_2+ 4 e_3
\frac{\pi'}{r}\right)}\,.
\ee
A non-trivial background profile for the scalar can change the $1/r^2$-dependence of the previous
expression  nearby the source. One easily check that the second term in the denominator
of the previous formula  becomes negligible when
 $r$ larger than a scale $r_A$,
given by
\be
r_A\,=\, \left[ \left(\hat{e}_3  \hat{d}_2\right)/\left(\hat{e}_2
\hat{d}_3 \right)\right]^{2/3}\,r_V\,.
\ee
 We can
 express this quantity using the formulae above, and find
\be
r_A\,=\,10^{10}\,  \left( \frac{\hat{e}_3^{\frac12}} {\hat{e}_2^{\frac23} \,\hat{d}_3^{\frac16}} \right)
\,\left(\frac{m_e}{\Lambda} \right)
\,r_{e}.
\ee
Hence we learn that $r_A$ is much larger than the electron radius, unless the coupling $\hat{e}_3$
is very small, or the scale $\Lambda$ well larger
than the electron mass. These simple considerations can
  lead to strong bounds on these
parameters,  and  show in a simple example that derivative
couplings between scalars with Galilean symmetry and electromagnetism can change considerably the behavior of electromagnetic and gravitational interactions.
In this section we focussed on the case of cubic Galileon interactions. A set-up  based on a quartic Galileon Lagrangian
will be analyzed in the Appendix.

\section{Summary}
Vector degrees of freedom typically arise in many examples of modified gravity models.
In this work, we started to systematically explore their role in these scenarios,
specifically studying the effects of derivatively coupled vectors and scalars.
To reduce the number of effective operators in the Lagrangian, we imposed appropriate symmetries
and physical constraints on the theory.  We required that our Lagrangian is invariant under a Galilean
symmetry in the scalar sector, and an Abelian gauge symmetry in the vector sector. Moreover,
in order to avoid  Ostrogradsky instabilities, we demanded that its associated equations of motion 
contain at most two space-time derivatives.

The resulting Lagrangian contains only a small number of terms. Starting from it, we investigated
the role of vector fields for two broad classes of phenomena that characterize modified gravity scenarios.
The first is self-acceleration. We analyzed in general terms the behavior of vector fluctuations
around self-accelerating solutions. We showed that it can provide key constraints to characterize instabilities
of cosmological backgrounds, in cases in which the kinetic terms of vector fluctuations vanish. 
 The second phenomenon we studied is the
screening of long range fifth forces in modified gravity models, in particular for the so-called Vainshtein mechanism.
In modified gravity scenarios based on Galileon symmetries, the non-linearities of field equations allow to screen 
the effects of light scalars within a certain distance (the Vainshtein radius) from a spherically symmetric source.
We showed that if a given source is characterized by a gravitational vector current, besides its usual energy 
momentum tensor, vectors play an important role for defining the background solution and the
scale corresponding to the Vainshtein radius. We also commented on how our findings can also be applied to set-ups 
in which scalars enjoying Galileon symmetries couple to bodies that are charged under electromagnetic interactions. 
Also in this case the realization of the Vainhstein mechanism might be influenced by the vector charges.

Our general results can be applied to any concrete model of modified gravity that satisfy the requirements 
that we imposed in this paper. 
It would be interesting to understand 
the dynamics of vectors also for space-times that are less symmetrical than the ones we considered.
For example, studying self-acceleration in cosmological space-times that break the isotropy of the 
Friedmann-Robertson-Walker Ansatz; or studying the realization of Vainshtein mechanism for
stationary space-times associated to sources that rotate around a given axis.  
 It can be expected that vector degrees of freedom can have interesting roles also in these set-ups,
 since they can acquire vacuum expectation values along a preferred spatial direction.

\subsection*{Acknowledgments}
We thank Gustavo Niz for useful comments on the draft. 
GT is supported by an STFC Advanced Fellowship ST/H005498/1.
KK is supported by STFC grant ST/H002774/1 and ST/K0090X/1, the European Research Council
and the Leverhulme trust.
NK acknowledges bilateral funding from the Royal Society and the
South African NRF which supported this project. NK also
thanks the Institute of Cosmology and Gravitation for its hospitality
during his visits.

\begin{appendix}

\section{The case of quartic galileon}

The case of quartic Galileon can be discussed similarly to the
cubic case. After substituting the vector equation,
the scalar equation to solve is
\be
d_2 \,y+2\,d_3\, y^2+2\,d_4\,y^3-\frac{\left(4 e_3 +2 e_4
y\right)\,Q_0^2}{r^{6}\,\left( 2 e_2+4 e_3 y+e_4
y^2\right)^2}\,=\,\frac{M}{M_{Pl}\, r^3}\label{geneomsc2}
\ee
In proximity of the source, we find the following solution for $y$
\bea
y &=&\frac{\hat{d}_2\,\Lambda^3}{\hat{d}_3}
\frac{r_V^{\pm}}{r}
\eea
with
\be
r_V^{\pm}\,\equiv\,\frac{\hat{d}_3}{\hat{d}_2\,\Lambda} \left[
\frac{M}{4\,\hat{d}_4\,M_{Pl}}\,\left( 1\pm\sqrt{1+\frac{16
\hat{d}_4 M_{Pl}^2 \,Q_0^2}{\hat{e}_4 M^2}} \right)\right]^{\frac13}
\ee
The Vainsthein radius corresponds to the scale at which
non-linear terms in the scalar equation become important. Also in
this case, a $Q_0\neq 0$ leads to two solutions for the Vainshtein
radius. As we will see, the requirement of
stability of the configuration imposes that the parameters
$\hat{d}_4$, $\hat{e}_4$ are positive: only the positive branch
$r_V^+$ is then allowed, and we will focus on it from now on. The
solution for the scalar and vector spherically symmetric
configuration is
\bea
\pi(r)&=& r_V\,r\\
A_0(r)&=&-\frac{Q_0 \,r}{e_4\,r_V}
\eea
The scalar contribution is much weaker than the usual gravitational one:
the Vainshtein mechanism is at work  and GR is recovered nearby a source.
Far from the source, instead, fifth forces become important.

\smallskip

Let us discuss also in this case the stability of these spherically
symmetric configurations. As before, we focus on spherically symmetric scalar and
vector perturbations, which respect the spherically symmetric Ansatz in terms of
the criteria used in \cite{Koyama:2011wx}:
\bea
\pi&=&\bar{\pi}(r)+\varphi(t,r)\,,\\
A_{\mu}&=&\left( \bar{A}_0(r)+\delta A_0(t,r),\,\delta
A_r(t,r),\,0,\,0\right)
\,,\\
F_{tr}&=&\delta \dot{A}_r(t,r)-\delta {A}_0'(t,r)\,.
\eea
The bars denote background quantities.

A de-mixing procedure of scalar from vectors requires a field redefinition
\be
\delta
F_{tr}\,\to\,-\frac{2\,\left(2\,e_3+e_4\,\bar{\pi}'\right)\,\bar{A}_0'\,\varphi'}
{\left(2e_2\,r+4\,e_3\,\bar{\pi}'+e_4\,\bar{\pi}'^2\right)}\,+\,\tilde{F}_{tr}\,,
\ee
hence we obtain the following Lagrangian for spherically symmetric
quadratic fluctuations around our spherically symmetric background
\be
S_{\varphi}\,=\,\frac12\,\int\,d^4 x\,\sqrt{-g}\,\left[
K_t(r)\,\dot{\varphi}^2-K_r(r)\,{\varphi'}^2-K_A(r)\,\tilde{F}_{tr}
\tilde{F}^{tr}\right]\,.
\ee
The functions  $K_i$ have to be positive in order to obtain a stable configuration. They read
\bea
K_t&=&\,{d_2}+4\,d_3\,\frac{\bar{\pi}'}{r}
+4\,d_4\,\frac{\bar{\pi}'^2}{r^2}
 +2\,d_3\,\bar{\pi}''
  +8\,d_4\,\frac{\bar{\pi}''\,\bar{\pi}'}{r}
 \,,
\\
K_r&=&\,
{d_2}+4\,d_3\,\frac{\bar{\pi}'}{r}+6\,d_4\,\frac{\bar{\pi}'^2}{r^2}
+\frac{ 2\,Q_0^2\,\left( 16 \,e_3^2-2 \,e_2\, e_4+12 \,e_3\,e_4\,
\bar{\pi}'/r+3 \,e_4^2 \,\bar{\pi}'^2/r^2\right)}{r^6\left(2 e_2+4
\, e_3 \bar{\pi}'/r +e_4 \,\bar{\pi}'^2/r^2 \right)^3} \,,
\\
K_A&=&
e_2+2\,e_3\,\frac{\bar{\pi}'}{r}+\frac{\,e_4}{2}\,\frac{\bar{\pi}'^2}{r^2}\,\,.
\eea
Nearby the source, the quantity $\bar{\pi}'/r$ is large: in
order to have positive $K_i$, we demand that $d_4$, $e_4$ are
positive. Far from the source, instead, $\bar{\pi}'/r$ is small: we
have to demand that  $d_2$, $e_2$ are positive. This implies that a
vector charge increases the size of the function $K_r$ above nearby
the source, while it gives only negligible contributions far from
it.

Computing the speed of radial scalar fluctuations in proximity of
the source we find
\be
c_r^2(r\to0)\,\equiv\,\frac{K_r}{K_t}(r\to0)\,=\, \frac23\,\left(
1+\frac{Q_0^2}{d_4\, e_4\,r_V^6}\right)
\ee
Hence the superluminal speed of radial fluctuations is increased by
the presence of the vector charge as in the cubic case.

\end{appendix}


\end{document}